\documentclass[prl,twocolumn,superscriptaddress,nobibnotes,a4paper]{revtex4-2}

\usepackage[pdftex]{graphicx}
\usepackage[pdftex]{epsfig}
\usepackage{amsmath}
\usepackage{amssymb}
\usepackage{amsfonts}
\usepackage{color}
\usepackage{wrapfig}
\usepackage{eucal}
\usepackage{hhline}
\usepackage{threeparttable}
\usepackage{supertabular}
\usepackage{multirow}
\usepackage{tabularx}
\usepackage[separate-uncertainty=true]{siunitx}
\usepackage{upgreek}
\usepackage{float}
\usepackage{siunitx}

\usepackage{mathtools}

\usepackage[export]{adjustbox}

\usepackage{soul}

\usepackage[version=4]{mhchem} % for chemical formulas
\usepackage{array} % make column bigger (vertical)
\usepackage[colorlinks=true, allcolors=blue]{hyperref} %hyperrefs

\usepackage{comment}

\setstcolor{red}

\begin{document}

\title{A cryogenic chamber setup for superfluid helium experiments with optical fiber and electrical access}\thanks{This work was published in \href{https://doi.org/10.1063/5.0249931}{Rev.\ Sci.\ Instrum.\ \textbf{96}, 083903 (2025).}}

\author{Alexander Rolf Korsch}
\affiliation{Department of Physics, Fudan University, Shanghai 200433, P.R.\ China}
\affiliation{Department of Physics, School of Science, Westlake University, Hangzhou 310030, P. R.\ China}
\affiliation{Kavli Institute of Nanoscience, Department of Quantum Nanoscience, Delft University of Technology, 2628CJ Delft, The Netherlands}
\author{Niccol\`{o} Fiaschi}
\affiliation{Kavli Institute of Nanoscience, Department of Quantum Nanoscience, Delft University of Technology, 2628CJ Delft, The Netherlands}
\author{Simon Gr\"oblacher}
\email{s.groeblacher@tudelft.nl}
\affiliation{Kavli Institute of Nanoscience, Department of Quantum Nanoscience, Delft University of Technology, 2628CJ Delft, The Netherlands}

%\date{\today}

\begin{abstract}
Superfluid helium is a prototypical quantum liquid. As such, it has been a prominent platform for the study of quantum many body physics. More recently, the outstanding mechanical and optical properties of superfluid helium, such as low mechanical dissipation and low optical absorption, have positioned superfluid helium as a promising material platform in applications ranging from dark matter and gravitational wave detection to quantum computation. However, experiments with superfluid helium incur a high barrier to entry as they require incorporation of complex optical and electrical setups within a hermetically sealed cryogenic chamber to confine the superfluid. Here, we report on the design and construction of a helium chamber setup for operation inside a dilution refrigerator at Millikelvin temperatures, featuring electrical and optical fiber access. By incorporating an automated gas handling system, we can precisely control the amount of helium gas inserted into the chamber, rendering our setup particularly promising for experiments with superfluid helium thin films, such as superfluid thin film optomechanics. Using silicon nanophotonic resonators, we demonstrate precise control and in-situ tuning of the thickness of a superfluid helium film on the sub-nanometer level. By making use of the exceptional tunability of the superfluid film thickness, we demonstrate optomechanically induced phonon lasing of phononic crystal cavity third sound modes in the superfluid film and show that the lasing threshold crucially depends on the film thickness.  The large internal volume of our chamber ($V_\mathrm{chamber} \approx \qty{1}{l}$) is adaptable for integration of various optical and electrical measurement and control techniques. Therefore, our setup provides a versatile platform for a variety of experiments in fundamental and applied superfluid helium research.
\end{abstract}

\newpage
\maketitle

\section*{Introduction}

Since its discovery in 1937 the experimental and theoretical study of superfluid helium has largely contributed to our understanding of macroscopic quantum phenomena. Up to this day, many fundamental questions about the physics of superfluid helium remain contentious and require further investigations, in particular in thin-films of superfluid governed by the dynamics of superfluid vortices. While such vortex dynamics have been studied extensively~\cite{kosterlitz_1973}, the exact dissipation mechanisms~\cite{adams_vortex_1987, thompson_quantum_2012} as well as the question of vortex inertia~\cite{thouless_vortex_2007, simula_vortex_2018} remain a topic of notable debate.

Besides its relevance for the study of fundamental quantum phenomena, superfluid helium has in recent years also become a subject of intense interest for applications in a wide range of research fields. Electrons floating on superfluid helium are being explored as a novel type of qubits for quantum computation due to their low-noise environment~\cite{kawakami_image-charge_2019, kawakami_blueprint_2023} and ability to couple to superconducting circuits for initialization and readout~\cite{yang_coupling_2016, koolstra_coupling_2019}. Furthermore, in the field of cavity optomechanics superfluid helium is receiving attention as a promising material platform for mechanical resonators due to its inherently low mechanical dissipation owing to its vanishingly small viscosity as well as its low optical absorption. Superfluid optomechanics has been realized both using bulk~\cite{lorenzo_superfluid_2014, kashkanova_superfluid_2017,shkarin_quantum_2019,patil_measuring_2022} and thin films of superfluid helium~\cite{harris_laser_2016}, enabling the study of superfluid vortex dynamics~\cite{sachkou_coherent_2019}, strong photothermal forces~\cite{sawadsky2023engineered}, and superfluid Brillouin lasing~\cite{he_strong_2020}. Control of mechanical excitations through phononic band structure engineering has been theoretically proposed in bulk superfluids~\cite{spence_superfluid_2021} and experimentally demonstrated in superfluid thin films~\cite{korsch_phononic_2024}. Superfluid optomechanics in laboratory-sized setups has been proposed for the detection of gravitational waves~\cite{singh_detecting_2017} as well as dark matter~\cite{hirschel_superfluid_2024, baker_optomechanical_2024}.

Experiments with superfluid helium thin films, in particular, require exceptional control over the thickness of the superfluid helium film. A particular example is the research field of superfluid thin film optomechanics~\cite{baker_theoretical_2016,harris_laser_2016, korsch_phononic_2024}:\ in these experiments the mechanical motion of the superfluid is studied by making use of its dispersive coupling to the optical mode of a nanophotonic resonator. The properties of the mechanical modes, such as their mechanical frequency and compliance, depend crucially on the thickness of the superfluid film. In addition to precise control over the helium film thickness, experiments in superfluid thin film optomechanics also require simultaneous optical and electrical control to allow for fiber coupling to the devices under test. Beyond current state-of-the-art experiments, exploring the functionalities of superfluid helium films in more elaborate applications will require the integration of complex experimental setups within a hermetically sealed environment suitable to confine superfluid helium, while at the same time allowing for precise control of the superfluid helium film thickness.

In recent years, a multitude of experimental setups for experiments with superfluid helium have been introduced for optical~\cite{kashkanova_phd} or microwave detection~\cite{de_lorenzo_optomechanics_2016} of bulk acoustic waves in superfluids. Moreover, setups compatible with superconducting circuits have been established to trap and control electrons floating on the surface of superfluid helium~\cite{koolstra_trapping_nodate}. In the field of superfluid thin film optomechanics, packaged experimental cells loaded with helium gas at room temperature and sealed with epoxy resin before cooldown have been developed, allowing robust plug-and-play functionality at the expense of in-situ tunability of the amount of helium gas filled into the chamber~\cite{wasserman_cryogenic_2022}. While few experiments report control over the helium film thickness~\cite{he_strong_2020}, a detailed description of a versatile platform for complex experiments involving optical and electrical access as well as exquisite control of the superfluid helium film thickness is still lacking.

Here, we present a complete guide on the design and construction of a helium chamber setup for use in a dilution refrigerator at Millikelvin temperatures meeting these requirements. Our chamber possesses a large volume of around 1 liter allowing for incorporation of large experimental setups. The large chamber volume also enables the incorporation of both electrical and optical access as well as piezo nanopositioners for optical coupling, allowing for complex experiments using both electrical and optical control techniques. An automated gas handling system enables precise control of the amount of helium gas inserted into the chamber and allows in-situ control of the film thickness of superfluid thin-films on the sub-nanometer level. To illustrate the capabilities of our setup, we demonstrate optomechanically induced phonon lasing of phononic crystal cavity third sound modes in a superfluid helium thin film on the surface of a silicon nanobeam photonic crystal resonator. Using the exceptional tunability of the superfluid helium film thickness, we verify a strong dependence of the phonon lasing threshold on the superfluid film thickness. In addition, although our setup is optimized for operation with superfluid thin films, we describe the procedure to operate in the bulk superfluid regime by filling the chamber with large quantities of superfluid.

\section*{Helium chamber design}

\begin{figure*}
	\centering
	\includegraphics[width = 18cm]{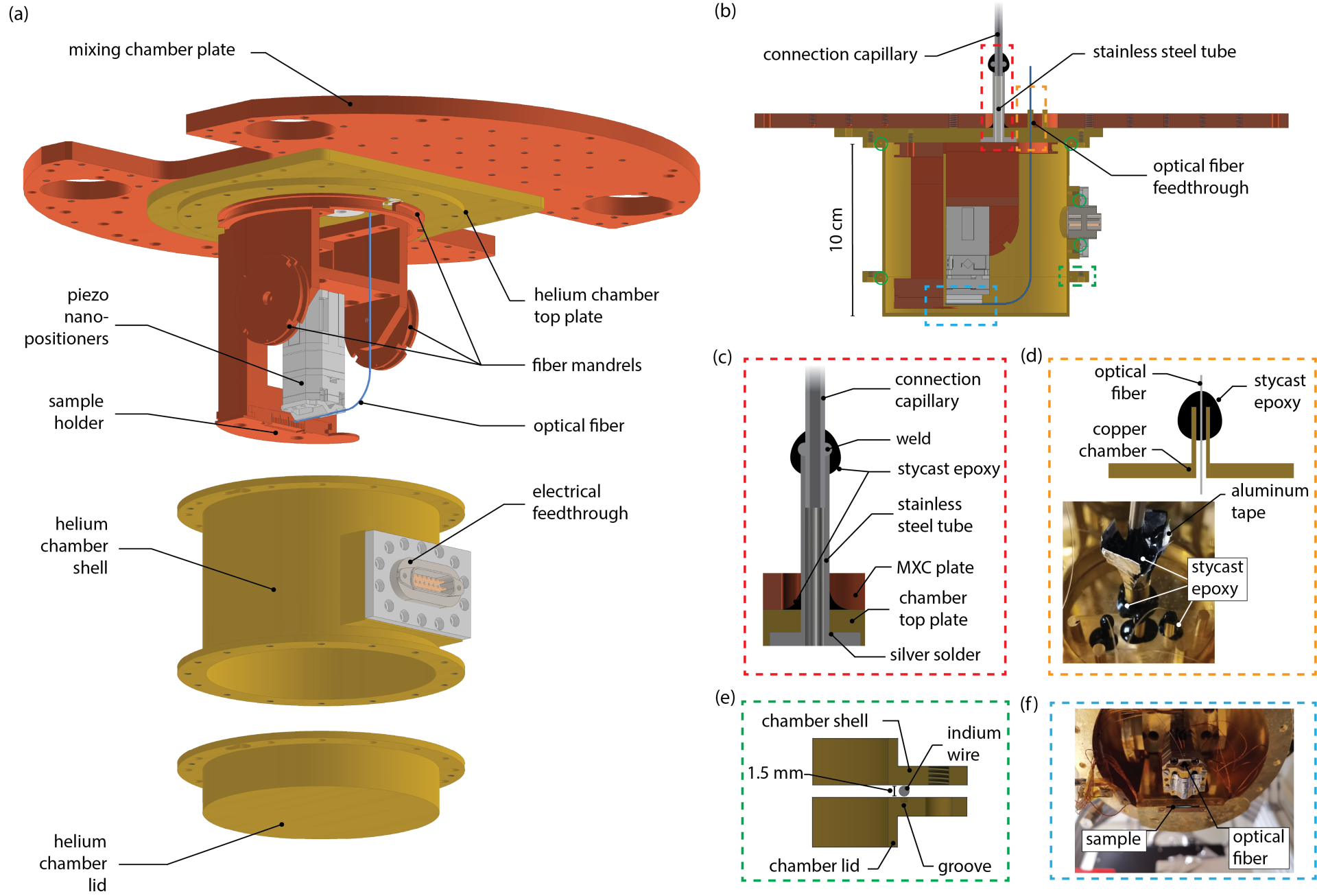}
	\caption{\textbf{Cryogenic helium chamber setup.} (a) Exploded sketch of the helium chamber setup mounted on the bottom of the mixing chamber (MXC) plate of a dilution refrigerator. (b) Projected half-section view from the side of the chamber when the shell and lid are closed. Optical fiber and sample coupling region (dashed blue box), locations of indium seals (green circles and dashed green box), optical fiber feedthroughs (dashed orange box), and connection to helium filling capillary (dashed red box) are highlighted and illustrated in more detail in (c), (d), (e), and (f), respectively. (c) Schematic illustration of the connection of the helium chamber to the helium filling capillary. (d) Top:\ illustration of the optical fiber feedthrough. Bottom:\ photograph of the optical fiber feedthrough and helium filling capillary on top of the chamber. (e) Illustration of indium wire seal between helium chamber shell and lid. (f) Photograph of the fiber and sample holder from the bottom when chamber lid is open.}
	\label{Fig:1_helium_chamber_setup}
\end{figure*}

Figures~\ref{Fig:1_helium_chamber_setup}(a) and (b) show an exploded drawing of our helium chamber setup and a projected view from the side, respectively. The chamber itself consists of three parts:\ a top plate that is designed to be mounted at the bottom of the mixing chamber plate of a dilution refrigerator, a shell covering the sides of the chamber, and a lid closing the bottom of the chamber. All three parts are milled from oxygen-free high-conductivity (OFHC) copper and the chamber top plate and shell are vacuum annealed for $\qty{48}{h}$ at a temperature of $\qty{840}{\degreeCelsius}$ to remove impurities and increase the thermal conductivity of the copper material~\cite{powell_lowtemperature_1959}. We use non-annealed OFHC copper for the helium chamber lid as we find that annealing the lid makes it too soft to withstand the pressure difference created by vacuum pumping leading to bending of the lid. To prevent surface oxidation and therefore reduce contact thermal resistance between different parts of the chamber, all parts are gold-plated after annealing. The chamber has a length of $\qty{10}{cm}$ and inner diameter of $\qty{10}{cm}$ with a wall thickness of $\qty{1.5}{mm}$.

Helium gas can be inserted into the chamber through a stainless steel filling capillary, which is connected to the chamber through the top plate (see Fig.~\ref{Fig:1_helium_chamber_setup}(c)). A $\qty{4}{cm}$ long piece of stainless steel tube with inner diameter of $\qty{3.4}{mm}$ is attached to the helium chamber top plate using silver-soldering. We then insert a $\qty{30}{cm}$ long connection piece of stainless steel capillary with outer diameter $\qty{3.175}{mm}$ into the stainless steel tube on top of the chamber. The two stainless steel tubes are welded together using tungsten inert gas welding. The silver soldering and welding procedures are perform before mounting the chamber top plate to the mixing chamber plate of the dilution refrigerator. To ensure leak-tight sealing, Stycast epoxy resin (2850 FT Loctite Stycast with 27-1 Loctite hardener) is used to seal the connection of the stainless steel connector to the chamber top plate. Due to the small size of the connecting tube, it is practically challenging to achieve leak-tight sealing of the welded connection between the stainless steel connector and the capillary. Therefore, we also cover the welded connection between the stainless steel tube and the connecting $\qty{30}{cm}$ long piece of stainless steel capillary in Stycast epoxy resin. To facilitate the deposition of the epoxy and ensure sufficient epoxy around the welded connection, we use aluminum tape to confine the epoxy while it is still liquid (see Fig.~\ref{Fig:1_helium_chamber_setup}(d)). After installation of the chamber top plate in the dilution refrigerator, the $\qty{30}{cm}$ long connection piece of stainless steel capillary can be manually bent to connect it to the rest of the helium filling capillary as discussed below. 

Optical fibers are inserted through milled cylindrical extension on top of the chamber top plate (see Fig.~\ref{Fig:1_helium_chamber_setup}(d)). These openings are then sealed using Stycast epoxy resin~\citep{pobell_matter_2007}. The cylindrically extended geometry of the fiber feedthrough allows improved sealing of the epoxy seal:\ during cooldown, the epoxy contracts more then the annealed copper material and thus compresses the annealed copper wall forming a reliable seal also at cryogenic temperature~\cite{he_micro_2020}. Copper-milled fiber mandrels oriented both vertically and horizontally are used inside the chamber to allow storage of several meters of optical fiber as well as orderly handling of the fiber inside the chamber. Storage of fiber inside the chamber allows splicing of lensed fibers or other fiber terminations without opening and sealing the epoxy fiber feedthrough. Both the connection to the helium filling capillary and the feedthrough of optical fibers are located at the top of the top plate of the helium chamber where the mixing chamber plate has a circular hole of diameter $\qty{40}{mm}$ allowing access to the chamber from the top.

To allow electrical access into the chamber, a 15-pin sub-D stainless steel connector (allectra 218-D15-SS) is welded to a stainless steel flange which is attached to the side of the helium chamber shell using an indium wire seal. The welding is performed by inserting the stainless steel connector into the opening in the flange and using the tungsten inert gas welding technique on the side of the flange facing outside of the chamber.

In order to hermetically seal the chamber against superfluid helium leakage, indium wire seals are used inbetween the parts of the chamber (see Fig.~\ref{Fig:1_helium_chamber_setup}(e)). The indium wire has a diameter of $\qty{1.5}{mm}$. The upper surface of the chamber shell, the chamber lid, and the stainless steel adapter for the electrical feedthrough feature a shallow groove of width $\qty{1.05}{mm}$ and depth $\qty{0.7}{mm}$ into which the indium wire can be inserted. By tightening the screws on each of the chamber parts ($16\times$ M4 shell to top plate, $16\times$ M4 lid to shell, $14\times$ M3 electrical feedthrough to shell), the indium wire is compressed, forming a cryogenically compatible vacuum and superfluid-tight seal. Before inserting the indium wire into the groove, we coat its surface with vacuum grease. The vacuum grease makes it easier to remove and exchange the indium wire when the chamber is opened again.

In principle, the large volume of our chamber setup $V_\mathrm{chamber}$ allows integration of a variety of different experimental setups. For our experiments, we utilize a suspended nanophotonic optical resonator that can be probed optically by lensed fiber coupling. Our sample is mounted on the bottom of a sample holder and the lensed optical fiber is mounted on a stack of three-axis piezo nanopositioners (Attocube ANPx101, ANPz102). The signals for control and readout of the nanopositioners are supplied through the electrical feedthrough of the chamber setup. A photograph of the device coupling region in our setup when the chamber lid is open is shown in Fig.~\ref{Fig:1_helium_chamber_setup}(f).

\section*{Gas handling system}

\begin{figure*}
	\centering
	\includegraphics[width = 18cm]{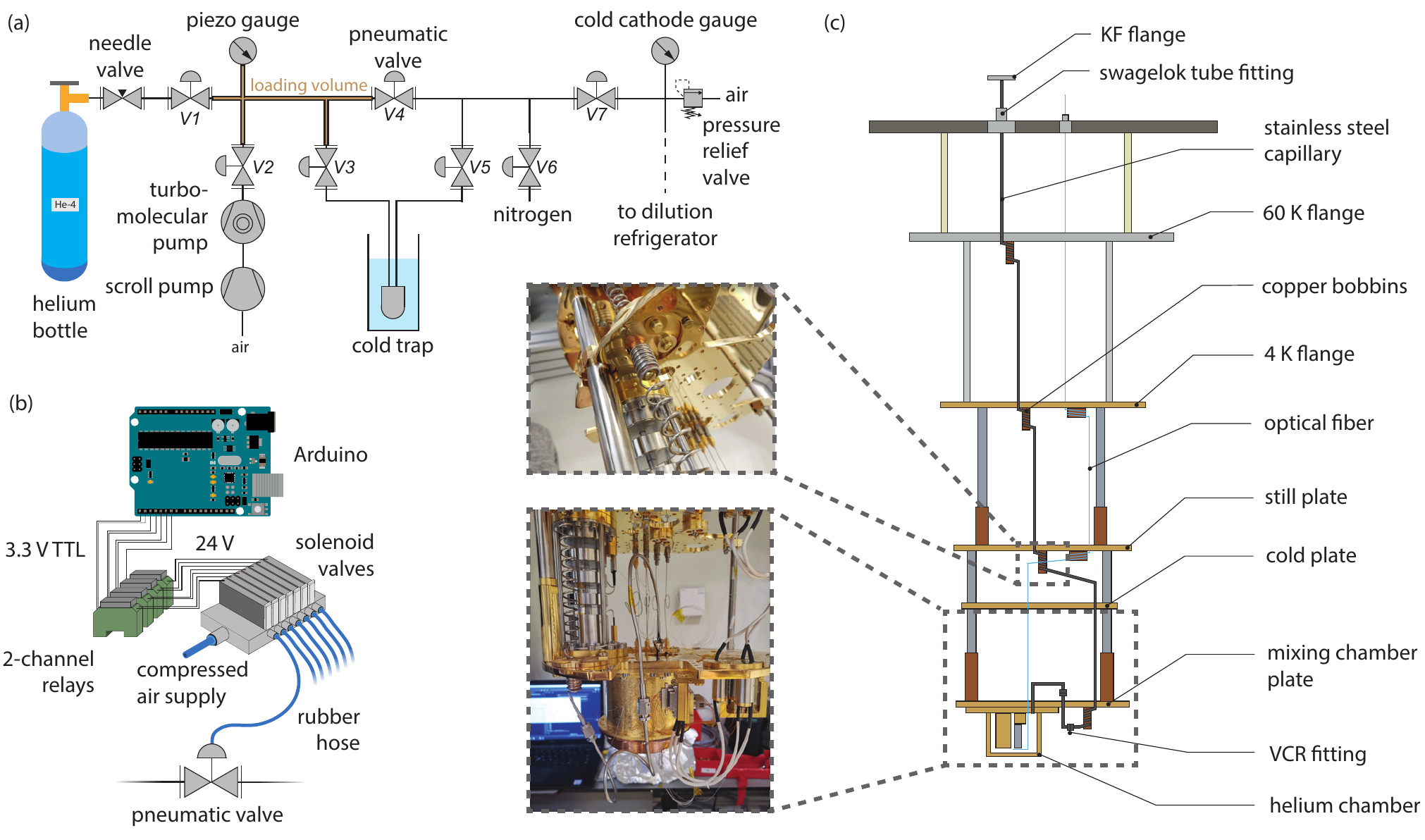}
	\caption{\textbf{Gas handling system.} (a) Sketch of the gas handling system outside of the dilution refrigerator used to load the helium chamber with helium-4 gas. (b) Schematic illustration of the control mechanism for the pneumatic valves. (c) Illustration of the connection of the helium filling capillary and optical fiber inside the dilution refrigerator. Insets in dashed boxes show photographs of the copper bobbins and connection of the helium filling capillary to the chamber.}
	\label{Fig:2_ghs}
\end{figure*}

To allow for precise and reproducible control of the amount of helium inserted into our chamber setup, we construct an automated gas handling system (see Fig.~\ref{Fig:2_ghs}(a)). The gas handling system utilizes seven pneumatic valves (Pfeiffer AIVP-S02100). Ultra-pure helium gas (99.9999\% elemental purity) is inserted from a bottle into the system through a dosing needle valve (Pfeiffer EVN 116). The gas handling system is connected to a scroll and turbomolecular pump (Pfeiffer HiCube 80 Pro) for vacuum pumping. The volume in between valves \textit{V1}, \textit{V2}, \textit{V3}, and \textit{V4} is used as a loading volume ($V_\mathrm{load} \approx \qty{0.5}{l}$) to set the amount of helium gas inserted into the chamber setup (see below). The helium gas pressure in the loading volume is measured using a piezo gauge (Pfeiffer Vacuum APR 265). To remove impurities, the gas in the loading volume can be expanded into an activated charcoal liquid nitrogen cold trap. The setup is further connected to a nitrogen supply line that can be used to flush both the gas handling system and the chamber setup. A pressure relief valve is located close to the connection of the gas handling system to the chamber setup in the dilution refrigerator to protect against unexpected pressure build-up in case of sudden evaporation of liquid helium in the chamber. A high-vacuum cold cathode pressure gauge (Pfeiffer PKR 360) is used to quantify the vacuum inside the gas handling system. The pneumatic valves used in the gas handling system are actuated through pressurized compressed air (see Fig.~\ref{Fig:2_ghs}(b)). For each pneumatic valve, an Arduino microcontroller switches a $\qty{24}{V}$ relay, which controls the compressed air supply ($p_\mathrm{air} = \qty{7}{bar}$) to the pneumatic valve via a solenoid valve (Festo VTUG series).

Figure~\ref{Fig:2_ghs}(c) shows a schematic illustration of the helium filling capillary and optical fiber connection inside the dilution refrigerator going from the top flange at room temperature to the mixing chamber plate at a base temperature of $\qty{10}{mK}$. The stainless steel filling capillary (length $l \approx \qty{3.7}{m}$, inner diameter $d=\qty{1.36}{mm}$) is welded to a KF-40 stainless steel flange that is mounted on one of the feedthroughs of our dilution refrigerator (Bluefors LD-250). On the outside of the dilution refrigerator, the flange is terminated with a swagelok tube fitting. To connect to the gas handling system, we utilize an adapter inserted into the tube fitting and terminated with a KF-16 flange. Inside the dilution refrigerator, the capillary is connected to the $\qty{60}{K}$ flange, $\qty{4}{K}$ flange, still plate, and mixing chamber plate by silver soldering to copper bobbins. At the mixing chamber plate, the filling capillary is terminated with a 1/8 inch VCR fitting to connect to the $\qty{30}{cm}$ long piece of stainless steel capillary welded to the top of the helium chamber top plate. Optical fibers are inserted into the dilution refrigerator using a well established feedthrough technique based on compressed teflon ferules~\cite{abraham_teflon_1998}. The optical fiber is spooled on copper spools at the $\qty{4}{K}$ flange and the still plate.

\section*{Experimental operation}

\begin{figure*}
	\centering
	\includegraphics[width = 18cm]{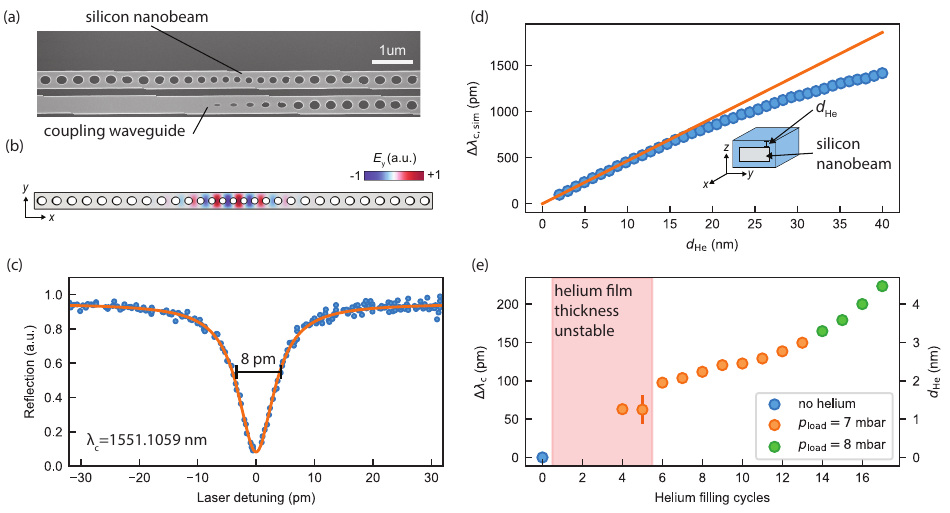}
	\caption{\textbf{Calibration of superfluid helium film thickness.} (a) Scanning electron microscope image of a suspended silicon nanobeam photonic crystal. (b) Finite-element simulation of the electric field component $E_y$ of the optical mode of the nanobeam (see Ref.~\cite{korsch_phononic_2024} for more details). (c) Optical reflection spectrum of the optical cavity resonance at resonance wavelength $\lambda_c = \qty{1551.1059}{nm}$ before any helium gas is inserted into the cavity. (d) Optical resonance wavelength shift $\Delta \lambda_\mathrm{c,sim}$ as simulated from finite element simulations as a function of superfluid helium film thickness $d_\mathrm{He}$ covering all surfaces of the suspended nanobeam resonator. The solid line is a linear extrapolation of the simulated data points for low film thickness $d_\mathrm{He} < \qty{10}{nm}$, yielding a linear shift of $\Delta \lambda_\mathrm{c,sim}/d_\mathrm{He}\approx 50$ picometer per nm film thickness~\cite{korsch_phononic_2024,baker_theoretical_2016}. (e) Experimentally measured shift of the optical cavity resonance $\Delta \lambda_\mathrm{c}$ as a function of helium filling cycles as described in the main text with loading volume pressure $p_\mathrm{load}$. Using the linear fit to the simulation result in (d), the experimentally measured resonance shift is converted to the helium film thickness. Red shaded area indicates the initial region when starting to fill the chamber with helium gas, in which the optical cavity resonance is not stable and could therefore not be measured precisely. Error bars are not visible for most data points as they are smaller than the marker size, except for in the region of unstable film thickness.}
	\label{Fig:3_helium_film_thickness}
\end{figure*}

We will now describe the detailed experimental operation procedure to fill the helium gas into the chamber setup. Furthermore, we describe our method to calibrate the thickness of superfluid helium thin films deposited this way utilizing suspended silicon photonic crystal cavity optical resonators. Lastly, we outline how we can use our setup to fill the chamber with large quantities of superfluid helium for bulk superfluid experiments.

\subsection*{Helium gas filling procedure}
\label{sec_filling_procedure}

Before cooldown of the cryostat, air in the cryogenic chamber setup is removed by flushing the chamber repeatedly with nitrogen and pumping it out. In the following, all valves of the gas handling system are assumed to be closed unless specified otherwise. For one flushing cycle, valve \textit{V6} is opened to fill the volume formed by valves \textit{V4}, \textit{V5}, \textit{V6}, and \textit{V7} with nitrogen gas at approximately $\qty{1}{bar}$ pressure. Then, \textit{V6} is closed and \textit{V7} is opened to expand the nitrogen gas into the filling capillary and chamber inside the cryostat. After a waiting time of $\qty{10}{min}$, valves \textit{V4} and \textit{V2} are opened to pump out the nitrogen gas from the setup. Pumping out is stopped by closing \textit{V2} once the pressure on the piezo gauge reaches $\qty{2}{mbar}$. We repeat this flushing cycle five times. After the last flushing cycle we leave valves \textit{V2}, \textit{V4}, and \textit{V7} open to evacuate the chamber for a longer time period of $\qty{24}{h}$. The flushing procedure is required to remove residual air in the chamber setup in the cryostat as the conductance of the filling capillary is very low. Therefore, vacuum pumping of the chamber does not fully remove air from the chamber: even after $\qty{24}{h}$ of vacuum pumping, the pressure in the chamber will only reach low vacuum levels. While we cannot measure the exact value of the pressure in the chamber directly, we estimate that we reach a pressure of $p_\mathrm{chamber} \approx \qty{0.1}{mbar}$ after $\qty{24}{h}$ of pumping based on the dimensions of the helium filling capillary. Afterwards, all valves are closed again. The remaining small amount of nitrogen gas freezes on the inner surfaces of the chamber when cooling down the cryostat. The presence of the thin nitrogen ice layer on the nanobeam surface leads to a red shift of the optical resonance by approximately $\qty{0.2}{nm}$, but it does not affect the linewidth of the optical cavity resonance.

Alternatively, the system can also be flushed using helium-4 gas instead of nitrogen to fully remove any non-helium contamination in the chamber. In this case, the remaining chamber pressure of helium-4 of $p_\mathrm{chamber} \approx \qty{0.1}{mbar}$ gas after $\qty{24}{h}$ of vacuum pumping, after cooldown to Millikelvin temperature, will lead to formation of an approximately  $\qty{20}{nm}$ thick helium film on all surfaces of the chamber. For experiments using nanometer-thin helium films as described below, this is undesirable and hence the flushing procedure is typically performed using nitrogen gas. We do not observe any quantifiable difference in the optical linewidth of the nanobeam resonator between flushing with nitrogen or helium gas.

To load helium gas into the chamber, during each loading cycle the loading volume (see Fig.~\ref{Fig:2_ghs}(a)) is filled with a predefined pressure $p_\mathrm{load}$, which can then be inserted through the filling capillary into the chamber setup. For this, the dosing needle valve is opened fully and the loading volume is first filled with approximately $\qty{1}{bar}$ of ultra-pure helium gas (99.9999\% elemental purity) by opening valve \textit{V1} for five seconds. To set the load volume pressure, the load volume is pumped out by opening valve \textit{V2} until the measured pressure on the piezo gauge reaches $p_\mathrm{load}$. The loaded helium gas is then expanded from the loading volume through the cold trap into the chamber setup by opening \textit{V3}, \textit{V5}, and \textit{V7} for one hour. We then close valve \textit{V7} and open \textit{V2} to pump out any helium gas remaining in the gas handling system. This method allows for arbitrary, precise and reproducible helium loading in the chamber.

\subsection*{Superfluid helium thin films}
\label{sec_thin_films}

Our chamber setup is particularly well suited for experiments using thin films of superfluid helium as it allows control of the helium film thickness on the sub-nanometer level. For such experiments, exquisite control over the superfluid film thickness is an essential prerequisite for systematic studies of superfluid film properties. A specific example is the research field of superfluid thin-film optomechanics:\ the superfluid helium film hosts surface acoustic waves -- so-called 'third sound' -- which can couple to the optical mode of optical micro- or nano-resonators by modulation of the effective refractive index of the confined optical mode. The restoring force of these third sound waves is given by the van der Waals acceleration between the superfluid and the optical resonator surface $g_\mathrm{vdW}=\alpha_\mathrm{vdW}/d_\mathrm{He}^4$, where $\alpha_\mathrm{vdW}$ is the Hamaker constant of the sample material and $d_\mathrm{He}$ is the thickness of the superfluid film. As a consequence, the mechanical properties of third sound waves, such as the mechanical frequency and compliance, depend strongly on the thickness of the superfluid film.

To generate thin films of superfluid helium, small quantities of helium gas are filled into the chamber. The helium gas thermalizes to the cryogenic environment, condenses and becomes superfluid. Due to its vanishing viscosity and attractive van der Waals forces to solids, superfluid helium forms a thin Rollin film that creeps up the walls of the chamber and therefore covers all exposed surfaces inside the chamber. Since the nanobeam resonator is suspended, the superfluid film also self-assembles on every surface of the nanobeam resonator. As more gas is inserted into the chamber, the superfluid film thickness increases up to a critical film thickness, which is determined by a balance between the superfluid-solid van der Waals force and gravity \cite{enss_low-temperature_2005}:
\begin{align}
\label{critical_film_thickness}
d_\mathrm{c} = \left(\frac{\alpha_\mathrm{vdW}}{g z_\mathrm{sample}}\right)^{1/3},
\end{align}
where $g=\qty{9.81}{m.s^{-2}}$ is the gravitational acceleration, and $z_\mathrm{sample}$ is the height of the sample with respect to the bottom of the chamber. Due to the thin layer of nitrogen ice on the surface of the nanobeam, the Hamaker constant is modified compared to the literature value for silicon and was determined as $\alpha_\mathrm{vdW}=\qty{0.9e-24}{m^5.s^{-2}}$ in a previous experiment~\cite{korsch_phononic_2024}. In our setup geometry, $z_\mathrm{sample} \approx \qty{3}{mm}$, yielding a critical film thickness $d_\mathrm{c} \approx \qty{30}{nm}$. The value of $d_\mathrm{c}$ puts a technical limit on the achievable superfluid film thickness:\ once the superfluid film thickness reaches the critical film thickness $d_\mathrm{c}$, excess superfluid gathers on the bottom of the chamber forming a bulk of superfluid instead of further increasing the thickness of the superfluid film. For experiments with superfluid thin films, we therefore operate our setup in the regime of unsaturated superfluid films, where $d_\mathrm{He}<d_\mathrm{c}$.

We demonstrate the control over the superfluid film thickness using a silicon nanobeam photonic crystal resonator (see Fig.~\ref{Fig:3_helium_film_thickness}(a)), which supports a high-quality optical mode in the telecom C-band wavelength range (see Fig.~\ref{Fig:3_helium_film_thickness}(b)). The formation of the superfluid helium film on the resonator surfaces leads to a red-shift of the optical resonator wavelength which we monitor by measuring the optical reflection spectrum of the cavity (see Fig.~\ref{Fig:3_helium_film_thickness}(c)). To allow for calibration of the helium film thickness from the observed resonance shift, we perform finite-element method simulations using COMSOL to simulate the optical resonance wavelength of our device as a function of helium film thickness $d_\mathrm{He}$ (see Fig.~\ref{Fig:3_helium_film_thickness}(d)). For very thin films where $d_\mathrm{He} < \qty{10}{nm}$, the optical resonance shift $\Delta \lambda_\mathrm{sim}$ depends linearly on the helium film thickness with a tuning of $\Delta \lambda_\mathrm{sim}/d_\mathrm{He} \approx \qty{50}{pm/nm}$.

Figure~\ref{Fig:3_helium_film_thickness}(e) shows the measured optical resonance shift $\lambda_\mathrm{c}$ as helium gas is let into the chamber in cycles of the filling procedure described above. The measured resonance shift is converted into the thickness of the helium film using the linear tuning parameter obtained from the simulation in Fig.~\ref{Fig:3_helium_film_thickness}(d). For the first five cycles of helium loading, the optical resonance is observed to be unstable. Repeated measurement of the optical resonance in this regime yields variations of the optical cavity resonance by more than the optical resonance linewidth. While the cause of this observation remains a subject for future investigation, we speculate that for very thin films close to the typical thickness of a superfluid helium monolayer $d_\mathrm{ml}\approx \qty{0.3}{nm}$~\cite{sabisky_verification_1973} the superfluid film might not be homogenous across the whole device. Theoretical studies have shown that in this regime of monolayer film thickness, the superfluid helium film tends to form patches instead of a homogenous film~\cite{clements_structure_1993, clements_growth_1993}. Heating of the film due to optical probing of the device may then lead to redistribution of the helium patches on the device surface leading to shifts and instability in the optical cavity resonance frequency. As more gas is inserted into the chamber, the helium film becomes stable and for $d_\mathrm{He}>\qty{2}{nm}$ the optical resonance can be measured reproducibly. By controlling the pressure of helium gas $p_\mathrm{load}$ loaded into the loading volume, we can control the change of helium film thickness per loading cycle. For $p_\mathrm{load}=\qty{7}{mbar}$, we increase the helium film thickness in increments of $\qty{0.15}{nm}$ per filling cycle, which is significantly smaller than the superfluid helium monolayer thickness $d_\mathrm{ml}\approx \qty{0.3}{nm}$~\cite{sabisky_verification_1973}. Owing to the narrow linewidth of the optical resonator mode $\kappa/2\pi = \qty{8}{pm}$ and considering the linear shift of the optical resonance as a function of helium thickness of $\qty{50}{pm/nm}$ our device can resolve a change in helium film thickness of $\Delta d_\mathrm{He} \approx \qty{0.16}{nm}$.

\begin{figure}
	\centering
	\includegraphics[width = 8.8cm]{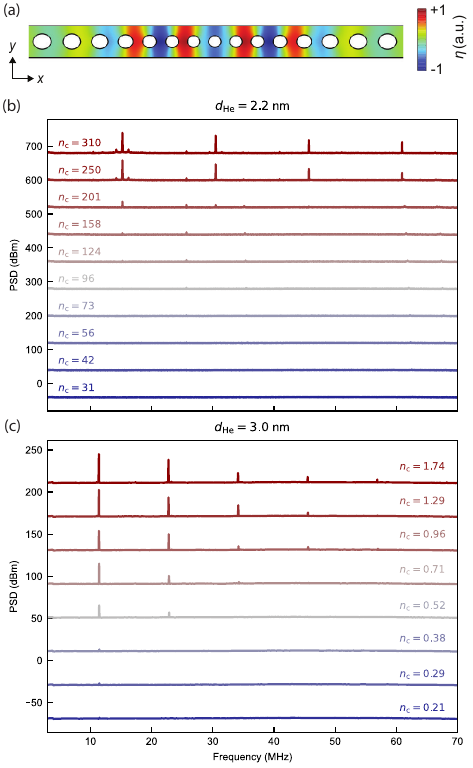}
	\caption{\textbf{Phonon lasing of phononic crystal third sound modes.} (a) Normalized variation $\eta$ of the superfluid film thickness for the fundamental third sound phononic crystal cavity mode in the superfluid helium film obtained from finite-element simulations. (b) Mechanical spectra of third sound phononic crystal cavity modes in the superfluid helium film measured using balanced homodyne detection at varying intracavity photon number $n_\mathrm{c}$ and helium film thickness $d_\mathrm{He}=\qty{2.2}{nm}$. (c) Mechanical spectra at varying intracavity photon number for a thicker helium film with $d_\mathrm{He}=\qty{3.0}{nm}$.}
	\label{Fig:4_phonon_lasing}
\end{figure}

As discussed in our previous work~\cite{korsch_phononic_2024}, the periodic patterning of the superfluid helium film on the surface of the nanobeam photonic crystal resonator gives rise to a third sound phononic band structure and enables confinement of third sound modes in phononic crystal cavities. The mode profile of the fundamental phononic crystal cavity mode is shown in Fig.~\ref{Fig:4_phonon_lasing}(a). The mechanical motion of these third sound modes can be probed through its optomechanical interaction with the optical cavity mode of the nanobeam. To demonstrate the importance of precise control over the superfluid film thickness for such experiments in superfluid thin film optomechanics, we measure the onset of phonon lasing of phononic crystal third sound modes for superfluid helium films with varying film thickness. Figure~\ref{Fig:4_phonon_lasing}(b) shows the mechanical spectra of the third sound modes measured via balanced homodyne detection at various intracavity photon number $n_\mathrm{c}$ and helium film thickness $d_\mathrm{He}=\qty{2.2}{nm}$. For $n_\mathrm{c} \geq 158$, we observe three peaks corresponding to different third sound modes and more modes become visible at increased intracavity photon number. At even higher intracavity photon numbers $n_\mathrm{c} \geq 250$, we observe higher harmonics of the fundamental third sound mode at frequency $\omega_\mathrm{m}/2\pi\approx\qty{15}{MHz}$. The observation of higher-order harmonics shows that in this regime the optomechanical interaction strongly drives the third sound motion through phonon lasing. The strong motional amplitude in turn leads to shifts of the optical cavity resonance on the order of the optical cavity linewidth $\kappa$. Such strong shifts of the optical cavity lead to nonlinear optomechanical transduction of the mechanical motion giving rise to higher order harmonics in the measured spectra.

For the thin film with $d_\mathrm{He}=\qty{2.2}{nm}$ in Fig.~\ref{Fig:4_phonon_lasing}(b), hundreds of intracavity photons are required to observe phonon lasing. However, a small increase of the superfluid film thickness to only $d_\mathrm{He}=\qty{3.0}{nm}$ leads to a significant reduction of the phonon lasing threshold and allows observation of higher harmonics for intracavity photon numbers $n_\mathrm{c}\approx 1$ as shown in Fig.~\ref{Fig:4_phonon_lasing}(c). The reduced phonon lasing threshold is a consequence of the reduced van der Waals acceleration $g_\mathrm{vdW}=\alpha_\mathrm{vdW}/d_\mathrm{He}^4$ for thicker films increasing the compliance of the superfluid film to surface deformations~\cite{sawadsky2023engineered}.

We note that for the measurements presented in Fig.~\ref{Fig:4_phonon_lasing} the laser is red-detuned by $\Delta/2\pi=\qty{-0.2}{GHz}$ with respect to the center of the optical cavity resonance. For conventional optomechanical systems, in which the optomechanical interaction is dominated by dispersive coupling, one would expect heating of the mechanical motion through optomechanical backaction and thus also phonon lasing to occur for blue-detuned laser driving, whereas red-detuned driving would lead to cooling. However, the optomechanical interaction for superfluid third sound modes is dominated by photothermal coupling, which can lead to an inverted sign of the optomechanical backaction~\cite{harris_laser_2016}. In particular, for the fundamental phononic crystal cavity third sound mode investigated in Fig.~\ref{Fig:4_phonon_lasing}, it has been shown that heating (cooling) of the mechanical motion occurs for red-detuned (blue-detuned) optical driving~\cite{korsch_phononic_2024}, consistent with the observation of phonon lasing for red-detuned driving.

\subsection*{Bulk superfluid helium}
\label{sec_bulk}

We note that for experiments using bulk superfluid helium our chamber setup can also be filled with large quantities of superfluid helium. Since the sample is mounted on a sample holder only $\qty{3}{mm}$ above the bottom of the chamber, the sample can be fully immersed in bulk superfluid by filling the chamber up to a filling level on the order of $\qty{3}{mm}$, corresponding to around $\qty{15}{cm^3}$ of liquid helium ($\qty{12}{l}$ of helium gas at standard pressure). Therefore, the geometry of our setup allows operation with minimal quantities of helium making it suitable also for experiments using bulk superfluid helium-4 or liquid helium-3, despite the large internal volume. Because the setup presented in this work mainly aims at experiments with superfluid thin films instead of bulk superfluid, the separation between the bottom of the chamber and the sample is chosen to be $\qty{3}{mm}$, to provide ample space for the piezo nanopositioners to move. However, this separation can easily be reduced to around $\qty{1}{mm}$ to further limit the amount of helium gas required.

To fill the chamber with large amounts of helium, we use a different filling procedure than described above by letting helium gas flow continuously into the chamber: We open \textit{V1}, \textit{V3}, \textit{V5}, and \textit{V7} and set the needle valve to control the gas flow to 0.1~mbar l/s. During this procedure, the presence of helium gas in the filling capillary creates a significant thermal link between the stages of the dilution refrigerator. Therefore, to avoid disrupting the helium mixture circulation when operating the dilution refrigerator at base temperature, we stop the circulation and operate the dilution refrigerator at a temperature of $\qty{4}{K}$. During filling, we continuously monitor the resonance wavelength of the optical resonator. Once the sample is immersed in bulk liquid helium, the optical resonance wavelength suddenly shifts by approximately $\qty{2}{nm}$ and stops shifting afterwards. For the settings of the needle valve specified above, reaching this point takes approximately $\qty{4}{h}$. Afterwards, we fill more helium gas into the chamber for another $\qty{1}{h}$ time period to ensure that the sample is fully immersed in liquid helium. After filling is completed, we cool down the dilution refrigerator to $\qty{10}{mK}$ base temperature so the filled liquid helium becomes superfluid.

\section*{Discussion and conclusion}

Superfluid helium is being explored as a promising material platform in different research areas of physics, ranging from dark matter and gravitational wave detection~\cite{baker_optomechanical_2024,hirschel_superfluid_2024} to quantum computation with electrons floating on the superfluid helium surface~\cite{kawakami_blueprint_2023}. Moreover, thin films of superfluid helium have been established as a promising material system for cavity optomechanics~\cite{baker_theoretical_2016}. Experiments with superfluid thin films, in particular, are challenging to perform, owing to the requirement of combining optical and electrical access within a cryogenic, superfluid-tight environment. Beyond current experiments, future hybrid approaches, combining multiple different quantum systems based on superfluid helium, will require accommodating even more electrical and optical measurement and control techniques.

Previous work on quantum experiments involving superfluid helium has mainly focused on experimental setups for experiments with bulk superfluid helium, including optical~\cite{kashkanova_optomechanics_2017,kashkanova_phd} and microwave~\cite{de_lorenzo_optomechanics_2016, lorenzo_superfluid_2014} probing of bulk acoustic waves. While the work by Koolstra \textit{et al.} on the trapping of electrons on the surface of bulk superfluid helium also introduces an automated gas handling system similar to the one used in our setup~\cite{koolstra_coupling_2019,koolstra_trapping_nodate}, their experiments neither demonstrate precise control over the superfluid film thickness on the sub-nm level nor incorporation of optical access. In the research field of superfluid thin film optomechanics, packaged experimental cells loaded with helium gas at room temperature and sealed with epoxy resist before cooldown have been introduced. Such setups offer robust operation and plug-and-play functionality, however, they do not allow in-situ tunability of the helium film thickness~\cite{wasserman_cryogenic_2022}. Few previous experiments demontrate control over the film thickness combined with optical access, for example by using alumina nanopowder to artificially increase the effective area of the chamber~\citep{he_strong_2020}. However, a full description of the experimental procedure to realize such tunability has not been reported.

This work presents the design and construction of a cryogenic helium chamber setup with fiber optical and electrical access meeting the stringent demands for experiments with superfluid helium thin films by incorporating optical and electrical access while allowing for sub-nm control of the superfluid film thickness. The chamber can be opened and closed using indium wire seals for hermetic sealing and reusable use. The gas handling system is designed to be fully automated, enabling the precise control of the helium filling procedure. The cryogenic chamber provides a large volume of around \qty{1}{l} that can accommodate a wide variety of experiments both in fundamental and applied research utilizing thin film superfluid helium. The large volume of our chamber as well as the availability of multiple feedthroughs for electrical and optical signals allows for integration of such complex experimental setups. In this work, we utilize these functionalities to enable lensed optical fiber coupling to nanophotonic devices using piezo nanopositioners. However, the same feedthrough techniques for optical and electrical access can in principle be used to enable various other experiments, such as insertion of RF signals into the chamber for experiments using superconducting circuits.

While our cryogenic chamber can also be filled with large quantities of helium gas to immerse samples in bulk superfluid, our experimental setup is particularly well suited for experiments with superfluid helium thin films as the automated gas handling system allows loading of the chamber with precisely dosed small quantities of helium gas. Using high-quality photonic crystal optical resonators, we demonstrate that we can control the thickness of superfluid helium films on the sub-nanometer level. Such exquisite control over the superfluid film thickness is crucial for enabling systematic studies of the dependence of physical phenomena in superfluid films on the film thickness. We demonstrate this by measuring the onset of optomechanically induced phonon lasing of confined third sound phononic crystal cavity modes in the superfluid film at two different values of the film thickness. Increasing the superfluid film thickness by only $\qty{0.8}{nm}$ from $\qty{2.2}{nm}$ to $\qty{3.0}{nm}$ reduces the threshold for phonon lasing by more than two orders of magnitude from an intracavity photon number $n_\mathrm{c}\approx 250$ to $n_\mathrm{c}\approx 1$. On the one hand, this observation illustrates the necessity of precise control over the superfluid film thickness. On the other hand, this is the first demonstration of phonon lasing for highly confined phononic crystal cavity third sound modes. Phonon lasing with very low lasing threshold around $n_\mathrm{c}\approx 1$ has previously only been demonstrated in whispering-gallery mode microsphere resonators, where it requires careful engineering of the device geometry and temperature to obtain strong photothermal coupling~\cite{sawadsky2023engineered}. Low-threshold phonon lasing enables excitation of third sound waves with large amplitudes without boiling of the superfluid film -- a fundamental requirement for studying superfluid hydrodynamics in the nonlinear regime~\cite{sawadsky2023engineered,kurihara_large-amplitude_1981,sachkou_coherent_2019}.

\medskip

\textbf{Acknowledgments}
We would like to thank Ronald Bode from the mechanical workshop at the Faculty of Applied Science at TU Delft for manufacturing parts of the helium chamber setup. We further acknowledge assistance from the Kavli Nanolab Delft. This work is financially supported by the European Research Council (ERC CoG Q-ECHOS, 101001005) and is part of the research program of the Netherlands Organization for Scientific Research (NWO), supported by the NWO Frontiers of Nanoscience program, as well as through a Vrij Programma (680-92-18-04) grant.

\textbf{Data Availability}
The data that support the findings of this study are available from the corresponding author upon reasonable request.


\begin{thebibliography}{37}%
	\makeatletter
	\providecommand \@ifxundefined [1]{%
		\@ifx{#1\undefined}
	}%
	\providecommand \@ifnum [1]{%
		\ifnum #1\expandafter \@firstoftwo
		\else \expandafter \@secondoftwo
		\fi
	}%
	\providecommand \@ifx [1]{%
		\ifx #1\expandafter \@firstoftwo
		\else \expandafter \@secondoftwo
		\fi
	}%
	\providecommand \natexlab [1]{#1}%
	\providecommand \enquote  [1]{``#1''}%
	\providecommand \bibnamefont  [1]{#1}%
	\providecommand \bibfnamefont [1]{#1}%
	\providecommand \citenamefont [1]{#1}%
	\providecommand \href@noop [0]{\@secondoftwo}%
	\providecommand \href [0]{\begingroup \@sanitize@url \@href}%
	\providecommand \@href[1]{\@@startlink{#1}\@@href}%
	\providecommand \@@href[1]{\endgroup#1\@@endlink}%
	\providecommand \@sanitize@url [0]{\catcode `\\12\catcode `\$12\catcode
		`\&12\catcode `\#12\catcode `\^12\catcode `\_12\catcode `\%12\relax}%
	\providecommand \@@startlink[1]{}%
	\providecommand \@@endlink[0]{}%
	\providecommand \url  [0]{\begingroup\@sanitize@url \@url }%
	\providecommand \@url [1]{\endgroup\@href {#1}{\urlprefix }}%
	\providecommand \urlprefix  [0]{URL }%
	\providecommand \Eprint [0]{\href }%
	\providecommand \doibase [0]{https://doi.org/}%
	\providecommand \selectlanguage [0]{\@gobble}%
	\providecommand \bibinfo  [0]{\@secondoftwo}%
	\providecommand \bibfield  [0]{\@secondoftwo}%
	\providecommand \translation [1]{[#1]}%
	\providecommand \BibitemOpen [0]{}%
	\providecommand \bibitemStop [0]{}%
	\providecommand \bibitemNoStop [0]{.\EOS\space}%
	\providecommand \EOS [0]{\spacefactor3000\relax}%
	\providecommand \BibitemShut  [1]{\csname bibitem#1\endcsname}%
	\let\auto@bib@innerbib\@empty
	%</preamble>
	\bibitem [{\citenamefont {Kosterlitz}\ and\ \citenamefont
		{Thouless}(1973)}]{kosterlitz_1973}%
	\BibitemOpen
	\bibfield  {author} {\bibinfo {author} {\bibfnamefont {J.~M.}\ \bibnamefont
			{Kosterlitz}}\ and\ \bibinfo {author} {\bibfnamefont {D.~J.}\ \bibnamefont
			{Thouless}},\ }\bibfield  {title} {\enquote {\bibinfo {title} {Ordering,
				metastability and phase transitions in two-dimensional systems},}\ }\href
	{https://doi.org/10.1088/0022-3719/6/7/010} {\bibfield  {journal} {\bibinfo
			{journal} {J. Phys. C: Solid State Phys.}\ }\textbf {\bibinfo {volume} {6}},\
		\bibinfo {pages} {1181} (\bibinfo {year} {1973})}\BibitemShut {NoStop}%
	\bibitem [{\citenamefont {Adams}\ and\ \citenamefont
		{Glaberson}(1987)}]{adams_vortex_1987}%
	\BibitemOpen
	\bibfield  {author} {\bibinfo {author} {\bibfnamefont {P.~W.}\ \bibnamefont
			{Adams}}\ and\ \bibinfo {author} {\bibfnamefont {W.~I.}\ \bibnamefont
			{Glaberson}},\ }\bibfield  {title} {\enquote {\bibinfo {title} {Vortex
				dynamics in superfluid helium films},}\ }\href
	{https://doi.org/10.1103/PhysRevB.35.4633} {\bibfield  {journal} {\bibinfo
			{journal} {Phys. Rev. B}\ }\textbf {\bibinfo {volume} {35}},\ \bibinfo
		{pages} {4633--4652} (\bibinfo {year} {1987})}\BibitemShut {NoStop}%
	\bibitem [{\citenamefont {Thompson}\ and\ \citenamefont
		{Stamp}(2012)}]{thompson_quantum_2012}%
	\BibitemOpen
	\bibfield  {author} {\bibinfo {author} {\bibfnamefont {L.}~\bibnamefont
			{Thompson}}\ and\ \bibinfo {author} {\bibfnamefont {P.~C.~E.}\ \bibnamefont
			{Stamp}},\ }\bibfield  {title} {\enquote {\bibinfo {title} {Quantum dynamics
				of a bose superfluid vortex},}\ }\href
	{https://doi.org/10.1103/PhysRevLett.108.184501} {\bibfield  {journal}
		{\bibinfo  {journal} {Phys. Rev. Lett.}\ }\textbf {\bibinfo {volume} {108}},\
		\bibinfo {pages} {184501} (\bibinfo {year} {2012})}\BibitemShut {NoStop}%
	\bibitem [{\citenamefont {Thouless}\ and\ \citenamefont
		{Anglin}(2007)}]{thouless_vortex_2007}%
	\BibitemOpen
	\bibfield  {author} {\bibinfo {author} {\bibfnamefont {D.~J.}\ \bibnamefont
			{Thouless}}\ and\ \bibinfo {author} {\bibfnamefont {J.~R.}\ \bibnamefont
			{Anglin}},\ }\bibfield  {title} {\enquote {\bibinfo {title} {Vortex mass in a
				superfluid at low frequencies},}\ }\href
	{https://doi.org/10.1103/PhysRevLett.99.105301} {\bibfield  {journal}
		{\bibinfo  {journal} {Phys. Rev. Lett.}\ }\textbf {\bibinfo {volume} {99}},\
		\bibinfo {pages} {105301} (\bibinfo {year} {2007})}\BibitemShut {NoStop}%
	\bibitem [{\citenamefont {Simula}(2018)}]{simula_vortex_2018}%
	\BibitemOpen
	\bibfield  {author} {\bibinfo {author} {\bibfnamefont {T.}~\bibnamefont
			{Simula}},\ }\bibfield  {title} {\enquote {\bibinfo {title} {Vortex mass in a
				superfluid},}\ }\href {https://doi.org/10.1103/PhysRevA.97.023609} {\bibfield
		{journal} {\bibinfo  {journal} {Phys. Rev. A}\ }\textbf {\bibinfo {volume}
			{97}},\ \bibinfo {pages} {023609} (\bibinfo {year} {2018})}\BibitemShut
	{NoStop}%
	\bibitem [{\citenamefont {Kawakami}, \citenamefont {Elarabi},\ and\
		\citenamefont {Konstantinov}(2019)}]{kawakami_image-charge_2019}%
	\BibitemOpen
	\bibfield  {author} {\bibinfo {author} {\bibfnamefont {E.}~\bibnamefont
			{Kawakami}}, \bibinfo {author} {\bibfnamefont {A.}~\bibnamefont {Elarabi}},\
		and\ \bibinfo {author} {\bibfnamefont {D.}~\bibnamefont {Konstantinov}},\
	}\bibfield  {title} {\enquote {\bibinfo {title} {Image-{Charge} {Detection}
				of the {Rydberg} {States} of {Surface} {Electrons} on {Liquid} {Helium}},}\
	}\href {https://doi.org/10.1103/PhysRevLett.123.086801} {\bibfield  {journal}
		{\bibinfo  {journal} {Phys. Rev. Lett.}\ }\textbf {\bibinfo {volume} {123}},\
		\bibinfo {pages} {086801} (\bibinfo {year} {2019})}\BibitemShut {NoStop}%
	\bibitem [{\citenamefont {Kawakami}\ \emph {et~al.}(2023)\citenamefont
		{Kawakami}, \citenamefont {Chen}, \citenamefont {Benito},\ and\ \citenamefont
		{Konstantinov}}]{kawakami_blueprint_2023}%
	\BibitemOpen
	\bibfield  {author} {\bibinfo {author} {\bibfnamefont {E.}~\bibnamefont
			{Kawakami}}, \bibinfo {author} {\bibfnamefont {J.}~\bibnamefont {Chen}},
		\bibinfo {author} {\bibfnamefont {M.}~\bibnamefont {Benito}},\ and\ \bibinfo
		{author} {\bibfnamefont {D.}~\bibnamefont {Konstantinov}},\ }\bibfield
	{title} {\enquote {\bibinfo {title} {Blueprint for quantum computing using
				electrons on helium},}\ }\href
	{https://doi.org/10.1103/PhysRevApplied.20.054022} {\bibfield  {journal}
		{\bibinfo  {journal} {Phys. Rev. Applied}\ }\textbf {\bibinfo {volume}
			{20}},\ \bibinfo {pages} {054022} (\bibinfo {year} {2023})}\BibitemShut
	{NoStop}%
	\bibitem [{\citenamefont {Yang}\ \emph {et~al.}(2016)\citenamefont {Yang},
		\citenamefont {Fragner}, \citenamefont {Koolstra}, \citenamefont {Ocola},
		\citenamefont {Czaplewski}, \citenamefont {Schoelkopf},\ and\ \citenamefont
		{Schuster}}]{yang_coupling_2016}%
	\BibitemOpen
	\bibfield  {author} {\bibinfo {author} {\bibfnamefont {G.}~\bibnamefont
			{Yang}}, \bibinfo {author} {\bibfnamefont {A.}~\bibnamefont {Fragner}},
		\bibinfo {author} {\bibfnamefont {G.}~\bibnamefont {Koolstra}}, \bibinfo
		{author} {\bibfnamefont {L.}~\bibnamefont {Ocola}}, \bibinfo {author}
		{\bibfnamefont {D.}~\bibnamefont {Czaplewski}}, \bibinfo {author}
		{\bibfnamefont {R.}~\bibnamefont {Schoelkopf}},\ and\ \bibinfo {author}
		{\bibfnamefont {D.}~\bibnamefont {Schuster}},\ }\bibfield  {title} {\enquote
		{\bibinfo {title} {Coupling an {Ensemble} of {Electrons} on {Superfluid}
				{Helium} to a {Superconducting} {Circuit}},}\ }\href
	{https://doi.org/10.1103/PhysRevX.6.011031} {\bibfield  {journal} {\bibinfo
			{journal} {Phys. Rev. X}\ }\textbf {\bibinfo {volume} {6}},\ \bibinfo {pages}
		{011031} (\bibinfo {year} {2016})}\BibitemShut {NoStop}%
	\bibitem [{\citenamefont {Koolstra}, \citenamefont {Yang},\ and\ \citenamefont
		{Schuster}(2019)}]{koolstra_coupling_2019}%
	\BibitemOpen
	\bibfield  {author} {\bibinfo {author} {\bibfnamefont {G.}~\bibnamefont
			{Koolstra}}, \bibinfo {author} {\bibfnamefont {G.}~\bibnamefont {Yang}},\
		and\ \bibinfo {author} {\bibfnamefont {D.~I.}\ \bibnamefont {Schuster}},\
	}\bibfield  {title} {\enquote {\bibinfo {title} {Coupling a single electron
				on superfluid helium to a superconducting resonator},}\ }\href
	{https://doi.org/10.1038/s41467-019-13335-7} {\bibfield  {journal} {\bibinfo
			{journal} {Nat. Commun.}\ }\textbf {\bibinfo {volume} {10}},\ \bibinfo
		{pages} {5323} (\bibinfo {year} {2019})}\BibitemShut {NoStop}%
	\bibitem [{\citenamefont {Lorenzo}\ and\ \citenamefont
		{Schwab}(2014)}]{lorenzo_superfluid_2014}%
	\BibitemOpen
	\bibfield  {author} {\bibinfo {author} {\bibfnamefont {L.~A.~D.}\
			\bibnamefont {Lorenzo}}\ and\ \bibinfo {author} {\bibfnamefont {K.~C.}\
			\bibnamefont {Schwab}},\ }\bibfield  {title} {\enquote {\bibinfo {title}
			{Superfluid optomechanics: coupling of a superfluid to a superconducting
				condensate},}\ }\href {https://doi.org/10.1088/1367-2630/16/11/113020}
	{\bibfield  {journal} {\bibinfo  {journal} {New J. Phys.}\ }\textbf {\bibinfo
			{volume} {16}},\ \bibinfo {pages} {113020} (\bibinfo {year}
		{2014})}\BibitemShut {NoStop}%
	\bibitem [{\citenamefont {Kashkanova}\ \emph
		{et~al.}(2017{\natexlab{a}})\citenamefont {Kashkanova}, \citenamefont
		{Shkarin}, \citenamefont {Brown}, \citenamefont {Flowers-Jacobs},
		\citenamefont {Childress}, \citenamefont {Hoch}, \citenamefont {Hohmann},
		\citenamefont {Ott}, \citenamefont {Reichel},\ and\ \citenamefont
		{Harris}}]{kashkanova_superfluid_2017}%
	\BibitemOpen
	\bibfield  {author} {\bibinfo {author} {\bibfnamefont {A.~D.}\ \bibnamefont
			{Kashkanova}}, \bibinfo {author} {\bibfnamefont {A.~B.}\ \bibnamefont
			{Shkarin}}, \bibinfo {author} {\bibfnamefont {C.~D.}\ \bibnamefont {Brown}},
		\bibinfo {author} {\bibfnamefont {N.~E.}\ \bibnamefont {Flowers-Jacobs}},
		\bibinfo {author} {\bibfnamefont {L.}~\bibnamefont {Childress}}, \bibinfo
		{author} {\bibfnamefont {S.~W.}\ \bibnamefont {Hoch}}, \bibinfo {author}
		{\bibfnamefont {L.}~\bibnamefont {Hohmann}}, \bibinfo {author} {\bibfnamefont
			{K.}~\bibnamefont {Ott}}, \bibinfo {author} {\bibfnamefont {J.}~\bibnamefont
			{Reichel}},\ and\ \bibinfo {author} {\bibfnamefont {J.~G.~E.}\ \bibnamefont
			{Harris}},\ }\bibfield  {title} {\enquote {\bibinfo {title} {Superfluid
				brillouin optomechanics},}\ }\href {https://doi.org/10.1038/nphys3900}
	{\bibfield  {journal} {\bibinfo  {journal} {Nat. Phys.}\ }\textbf {\bibinfo
			{volume} {13}},\ \bibinfo {pages} {74--79} (\bibinfo {year}
		{2017}{\natexlab{a}})}\BibitemShut {NoStop}%
	\bibitem [{\citenamefont {Shkarin}\ \emph {et~al.}(2019)\citenamefont
		{Shkarin}, \citenamefont {Kashkanova}, \citenamefont {Brown}, \citenamefont
		{Garcia}, \citenamefont {Ott}, \citenamefont {Reichel},\ and\ \citenamefont
		{Harris}}]{shkarin_quantum_2019}%
	\BibitemOpen
	\bibfield  {author} {\bibinfo {author} {\bibfnamefont {A.}~\bibnamefont
			{Shkarin}}, \bibinfo {author} {\bibfnamefont {A.}~\bibnamefont {Kashkanova}},
		\bibinfo {author} {\bibfnamefont {C.}~\bibnamefont {Brown}}, \bibinfo
		{author} {\bibfnamefont {S.}~\bibnamefont {Garcia}}, \bibinfo {author}
		{\bibfnamefont {K.}~\bibnamefont {Ott}}, \bibinfo {author} {\bibfnamefont
			{J.}~\bibnamefont {Reichel}},\ and\ \bibinfo {author} {\bibfnamefont
			{J.}~\bibnamefont {Harris}},\ }\bibfield  {title} {\enquote {\bibinfo {title}
			{Quantum {Optomechanics} in a {Liquid}},}\ }\href
	{https://doi.org/10.1103/PhysRevLett.122.153601} {\bibfield  {journal}
		{\bibinfo  {journal} {Phys. Rev. Lett.}\ }\textbf {\bibinfo {volume} {122}},\
		\bibinfo {pages} {153601} (\bibinfo {year} {2019})}\BibitemShut {NoStop}%
	\bibitem [{\citenamefont {Patil}\ \emph {et~al.}(2022)\citenamefont {Patil},
		\citenamefont {Yu}, \citenamefont {Frazier}, \citenamefont {Wang},
		\citenamefont {Johnson}, \citenamefont {Fox}, \citenamefont {Reichel},\ and\
		\citenamefont {Harris}}]{patil_measuring_2022}%
	\BibitemOpen
	\bibfield  {author} {\bibinfo {author} {\bibfnamefont {Y.~S.~S.}\
			\bibnamefont {Patil}}, \bibinfo {author} {\bibfnamefont {J.}~\bibnamefont
			{Yu}}, \bibinfo {author} {\bibfnamefont {S.}~\bibnamefont {Frazier}},
		\bibinfo {author} {\bibfnamefont {Y.}~\bibnamefont {Wang}}, \bibinfo {author}
		{\bibfnamefont {K.}~\bibnamefont {Johnson}}, \bibinfo {author} {\bibfnamefont
			{J.}~\bibnamefont {Fox}}, \bibinfo {author} {\bibfnamefont {J.}~\bibnamefont
			{Reichel}},\ and\ \bibinfo {author} {\bibfnamefont {J.~G.~E.}\ \bibnamefont
			{Harris}},\ }\bibfield  {title} {\enquote {\bibinfo {title} {Measuring
				high-order phonon correlations in an optomechanical resonator},}\ }\href
	{https://doi.org/10.1103/PhysRevLett.128.183601} {\bibfield  {journal}
		{\bibinfo  {journal} {Phys. Rev. Lett.}\ }\textbf {\bibinfo {volume} {128}},\
		\bibinfo {pages} {183601} (\bibinfo {year} {2022})}\BibitemShut {NoStop}%
	\bibitem [{\citenamefont {Harris}\ \emph {et~al.}(2016)\citenamefont {Harris},
		\citenamefont {McAuslan}, \citenamefont {Sheridan}, \citenamefont {Sachkou},
		\citenamefont {Baker},\ and\ \citenamefont {Bowen}}]{harris_laser_2016}%
	\BibitemOpen
	\bibfield  {author} {\bibinfo {author} {\bibfnamefont {G.~I.}\ \bibnamefont
			{Harris}}, \bibinfo {author} {\bibfnamefont {D.~L.}\ \bibnamefont
			{McAuslan}}, \bibinfo {author} {\bibfnamefont {E.}~\bibnamefont {Sheridan}},
		\bibinfo {author} {\bibfnamefont {Y.}~\bibnamefont {Sachkou}}, \bibinfo
		{author} {\bibfnamefont {C.}~\bibnamefont {Baker}},\ and\ \bibinfo {author}
		{\bibfnamefont {W.~P.}\ \bibnamefont {Bowen}},\ }\bibfield  {title} {\enquote
		{\bibinfo {title} {Laser cooling and control of excitations in superfluid
				helium},}\ }\href {https://doi.org/10.1038/nphys3714} {\bibfield  {journal}
		{\bibinfo  {journal} {Nat. Phys.}\ }\textbf {\bibinfo {volume} {12}},\
		\bibinfo {pages} {788--793} (\bibinfo {year} {2016})}\BibitemShut {NoStop}%
	\bibitem [{\citenamefont {Sachkou}\ \emph {et~al.}(2019)\citenamefont
		{Sachkou}, \citenamefont {Baker}, \citenamefont {Harris}, \citenamefont
		{Stockdale}, \citenamefont {Forstner}, \citenamefont {Reeves}, \citenamefont
		{He}, \citenamefont {McAuslan}, \citenamefont {Bradley}, \citenamefont
		{Davis},\ and\ \citenamefont {Bowen}}]{sachkou_coherent_2019}%
	\BibitemOpen
	\bibfield  {author} {\bibinfo {author} {\bibfnamefont {Y.~P.}\ \bibnamefont
			{Sachkou}}, \bibinfo {author} {\bibfnamefont {C.~G.}\ \bibnamefont {Baker}},
		\bibinfo {author} {\bibfnamefont {G.~I.}\ \bibnamefont {Harris}}, \bibinfo
		{author} {\bibfnamefont {O.~R.}\ \bibnamefont {Stockdale}}, \bibinfo {author}
		{\bibfnamefont {S.}~\bibnamefont {Forstner}}, \bibinfo {author}
		{\bibfnamefont {M.~T.}\ \bibnamefont {Reeves}}, \bibinfo {author}
		{\bibfnamefont {X.}~\bibnamefont {He}}, \bibinfo {author} {\bibfnamefont
			{D.~L.}\ \bibnamefont {McAuslan}}, \bibinfo {author} {\bibfnamefont {A.~S.}\
			\bibnamefont {Bradley}}, \bibinfo {author} {\bibfnamefont {M.~J.}\
			\bibnamefont {Davis}},\ and\ \bibinfo {author} {\bibfnamefont {W.~P.}\
			\bibnamefont {Bowen}},\ }\bibfield  {title} {\enquote {\bibinfo {title}
			{Coherent vortex dynamics in a strongly interacting superfluid on a silicon
				chip},}\ }\href {https://doi.org/10.1126/science.aaw9229} {\bibfield
		{journal} {\bibinfo  {journal} {Science}\ }\textbf {\bibinfo {volume}
			{366}},\ \bibinfo {pages} {1480--1485} (\bibinfo {year} {2019})}\BibitemShut
	{NoStop}%
	\bibitem [{\citenamefont {Sawadsky}\ \emph {et~al.}(2023)\citenamefont
		{Sawadsky}, \citenamefont {Harrison}, \citenamefont {Harris}, \citenamefont
		{Wasserman}, \citenamefont {Sfendla}, \citenamefont {Bowen},\ and\
		\citenamefont {Baker}}]{sawadsky2023engineered}%
	\BibitemOpen
	\bibfield  {author} {\bibinfo {author} {\bibfnamefont {A.}~\bibnamefont
			{Sawadsky}}, \bibinfo {author} {\bibfnamefont {R.~A.}\ \bibnamefont
			{Harrison}}, \bibinfo {author} {\bibfnamefont {G.~I.}\ \bibnamefont
			{Harris}}, \bibinfo {author} {\bibfnamefont {W.~W.}\ \bibnamefont
			{Wasserman}}, \bibinfo {author} {\bibfnamefont {Y.~L.}\ \bibnamefont
			{Sfendla}}, \bibinfo {author} {\bibfnamefont {W.~P.}\ \bibnamefont {Bowen}},\
		and\ \bibinfo {author} {\bibfnamefont {C.~G.}\ \bibnamefont {Baker}},\
	}\bibfield  {title} {\enquote {\bibinfo {title} {Engineered entropic forces
				allow ultrastrong dynamical backaction},}\ }\href
	{https://doi.org/10.1126/sciadv.ade3591} {\bibfield  {journal} {\bibinfo
			{journal} {Sci. Adv.}\ }\textbf {\bibinfo {volume} {9}},\ \bibinfo {pages}
		{eade3591} (\bibinfo {year} {2023})}\BibitemShut {NoStop}%
	\bibitem [{\citenamefont {He}\ \emph {et~al.}(2020)\citenamefont {He},
		\citenamefont {Harris}, \citenamefont {Baker}, \citenamefont {Sawadsky},
		\citenamefont {Sfendla}, \citenamefont {Sachkou}, \citenamefont {Forstner},\
		and\ \citenamefont {Bowen}}]{he_strong_2020}%
	\BibitemOpen
	\bibfield  {author} {\bibinfo {author} {\bibfnamefont {X.}~\bibnamefont
			{He}}, \bibinfo {author} {\bibfnamefont {G.~I.}\ \bibnamefont {Harris}},
		\bibinfo {author} {\bibfnamefont {C.~G.}\ \bibnamefont {Baker}}, \bibinfo
		{author} {\bibfnamefont {A.}~\bibnamefont {Sawadsky}}, \bibinfo {author}
		{\bibfnamefont {Y.~L.}\ \bibnamefont {Sfendla}}, \bibinfo {author}
		{\bibfnamefont {Y.~P.}\ \bibnamefont {Sachkou}}, \bibinfo {author}
		{\bibfnamefont {S.}~\bibnamefont {Forstner}},\ and\ \bibinfo {author}
		{\bibfnamefont {W.~P.}\ \bibnamefont {Bowen}},\ }\bibfield  {title} {\enquote
		{\bibinfo {title} {Strong optical coupling through superfluid {Brillouin}
				lasing},}\ }\href {https://doi.org/10.1038/s41567-020-0785-0} {\bibfield
		{journal} {\bibinfo  {journal} {Nat. Phys.}\ }\textbf {\bibinfo {volume}
			{16}},\ \bibinfo {pages} {417--421} (\bibinfo {year} {2020})}\BibitemShut
	{NoStop}%
	\bibitem [{\citenamefont {Spence}\ \emph {et~al.}(2021)\citenamefont {Spence},
		\citenamefont {Koong}, \citenamefont {Horsley},\ and\ \citenamefont
		{Rojas}}]{spence_superfluid_2021}%
	\BibitemOpen
	\bibfield  {author} {\bibinfo {author} {\bibfnamefont {S.}~\bibnamefont
			{Spence}}, \bibinfo {author} {\bibfnamefont {Z.}~\bibnamefont {Koong}},
		\bibinfo {author} {\bibfnamefont {S.}~\bibnamefont {Horsley}},\ and\ \bibinfo
		{author} {\bibfnamefont {X.}~\bibnamefont {Rojas}},\ }\bibfield  {title}
	{\enquote {\bibinfo {title} {Superfluid {Optomechanics} {With} {Phononic}
				{Nanostructures}},}\ }\href
	{https://doi.org/10.1103/PhysRevApplied.15.034090} {\bibfield  {journal}
		{\bibinfo  {journal} {Phys. Rev. Applied}\ }\textbf {\bibinfo {volume}
			{15}},\ \bibinfo {pages} {034090} (\bibinfo {year} {2021})}\BibitemShut
	{NoStop}%
	\bibitem [{\citenamefont {Korsch}, \citenamefont {Fiaschi},\ and\ \citenamefont
		{Gr\"oblacher}(2024)}]{korsch_phononic_2024}%
	\BibitemOpen
	\bibfield  {author} {\bibinfo {author} {\bibfnamefont {A.~R.}\ \bibnamefont
			{Korsch}}, \bibinfo {author} {\bibfnamefont {N.}~\bibnamefont {Fiaschi}},\
		and\ \bibinfo {author} {\bibfnamefont {S.}~\bibnamefont {Gr\"oblacher}},\
	}\bibfield  {title} {\enquote {\bibinfo {title} {Phononic crystals in
				superfluid thin-film helium},}\ }\href
	{https://doi.org/10.1103/PhysRevApplied.22.L041005} {\bibfield  {journal}
		{\bibinfo  {journal} {Phys. Rev. Applied}\ }\textbf {\bibinfo {volume}
			{22}},\ \bibinfo {pages} {L041005} (\bibinfo {year} {2024})}\BibitemShut
	{NoStop}%
	\bibitem [{\citenamefont {Singh}\ \emph {et~al.}(2017)\citenamefont {Singh},
		\citenamefont {Lorenzo}, \citenamefont {Pikovski},\ and\ \citenamefont
		{Schwab}}]{singh_detecting_2017}%
	\BibitemOpen
	\bibfield  {author} {\bibinfo {author} {\bibfnamefont {S.}~\bibnamefont
			{Singh}}, \bibinfo {author} {\bibfnamefont {L.~A.~D.}\ \bibnamefont
			{Lorenzo}}, \bibinfo {author} {\bibfnamefont {I.}~\bibnamefont {Pikovski}},\
		and\ \bibinfo {author} {\bibfnamefont {K.~C.}\ \bibnamefont {Schwab}},\
	}\bibfield  {title} {\enquote {\bibinfo {title} {Detecting continuous
				gravitational waves with superfluid {4He}},}\ }\href
	{https://doi.org/10.1088/1367-2630/aa78cb} {\bibfield  {journal} {\bibinfo
			{journal} {New J. Phys.}\ }\textbf {\bibinfo {volume} {19}},\ \bibinfo
		{pages} {073023} (\bibinfo {year} {2017})}\BibitemShut {NoStop}%
	\bibitem [{\citenamefont {Hirschel}\ \emph {et~al.}(2024)\citenamefont
		{Hirschel}, \citenamefont {Vadakkumbatt}, \citenamefont {Baker},
		\citenamefont {Schweizer}, \citenamefont {Sankey}, \citenamefont {Singh},\
		and\ \citenamefont {Davis}}]{hirschel_superfluid_2024}%
	\BibitemOpen
	\bibfield  {author} {\bibinfo {author} {\bibfnamefont {M.}~\bibnamefont
			{Hirschel}}, \bibinfo {author} {\bibfnamefont {V.}~\bibnamefont
			{Vadakkumbatt}}, \bibinfo {author} {\bibfnamefont {N.}~\bibnamefont {Baker}},
		\bibinfo {author} {\bibfnamefont {F.}~\bibnamefont {Schweizer}}, \bibinfo
		{author} {\bibfnamefont {J.}~\bibnamefont {Sankey}}, \bibinfo {author}
		{\bibfnamefont {S.}~\bibnamefont {Singh}},\ and\ \bibinfo {author}
		{\bibfnamefont {J.}~\bibnamefont {Davis}},\ }\bibfield  {title} {\enquote
		{\bibinfo {title} {Superfluid helium ultralight dark matter detector},}\
	}\href {https://doi.org/10.1103/PhysRevD.109.095011} {\bibfield  {journal}
		{\bibinfo  {journal} {Phys. Rev. D}\ }\textbf {\bibinfo {volume} {109}},\
		\bibinfo {pages} {095011} (\bibinfo {year} {2024})}\BibitemShut {NoStop}%
	\bibitem [{\citenamefont {Baker}\ \emph {et~al.}(2024)\citenamefont {Baker},
		\citenamefont {Bowen}, \citenamefont {Cox}, \citenamefont {Dolan},
		\citenamefont {Goryachev},\ and\ \citenamefont
		{Harris}}]{baker_optomechanical_2024}%
	\BibitemOpen
	\bibfield  {author} {\bibinfo {author} {\bibfnamefont {C.~G.}\ \bibnamefont
			{Baker}}, \bibinfo {author} {\bibfnamefont {W.~P.}\ \bibnamefont {Bowen}},
		\bibinfo {author} {\bibfnamefont {P.}~\bibnamefont {Cox}}, \bibinfo {author}
		{\bibfnamefont {M.~J.}\ \bibnamefont {Dolan}}, \bibinfo {author}
		{\bibfnamefont {M.}~\bibnamefont {Goryachev}},\ and\ \bibinfo {author}
		{\bibfnamefont {G.}~\bibnamefont {Harris}},\ }\bibfield  {title} {\enquote
		{\bibinfo {title} {Optomechanical dark matter instrument for direct
				detection},}\ }\href {https://doi.org/10.1103/PhysRevD.110.043005} {\bibfield
		{journal} {\bibinfo  {journal} {Phys. Rev. D}\ }\textbf {\bibinfo {volume}
			{110}},\ \bibinfo {pages} {043005} (\bibinfo {year} {2024})}\BibitemShut
	{NoStop}%
	\bibitem [{\citenamefont {Baker}\ \emph {et~al.}(2016)\citenamefont {Baker},
		\citenamefont {Harris}, \citenamefont {McAuslan}, \citenamefont {Sachkou},
		\citenamefont {He},\ and\ \citenamefont {Bowen}}]{baker_theoretical_2016}%
	\BibitemOpen
	\bibfield  {author} {\bibinfo {author} {\bibfnamefont {C.~G.}\ \bibnamefont
			{Baker}}, \bibinfo {author} {\bibfnamefont {G.~I.}\ \bibnamefont {Harris}},
		\bibinfo {author} {\bibfnamefont {D.~L.}\ \bibnamefont {McAuslan}}, \bibinfo
		{author} {\bibfnamefont {Y.}~\bibnamefont {Sachkou}}, \bibinfo {author}
		{\bibfnamefont {X.}~\bibnamefont {He}},\ and\ \bibinfo {author}
		{\bibfnamefont {W.~P.}\ \bibnamefont {Bowen}},\ }\bibfield  {title} {\enquote
		{\bibinfo {title} {Theoretical framework for thin film superfluid
				optomechanics: towards the quantum regime},}\ }\href
	{https://doi.org/10.1088/1367-2630/aa520d} {\bibfield  {journal} {\bibinfo
			{journal} {New J. Phys.}\ }\textbf {\bibinfo {volume} {18}},\ \bibinfo
		{pages} {123025} (\bibinfo {year} {2016})}\BibitemShut {NoStop}%
	\bibitem [{\citenamefont {Kashkanova}(2018)}]{kashkanova_phd}%
	\BibitemOpen
	\bibfield  {author} {\bibinfo {author} {\bibfnamefont {A.}~\bibnamefont
			{Kashkanova}},\ }\emph {\bibinfo {title} {Optomechanics with {Superfluid}
			{Helium}}},\ \href@noop {} {Ph.D. thesis},\ \bibinfo  {school} {The
		University of Queensland} (\bibinfo {year} {2018})\BibitemShut {NoStop}%
	\bibitem [{\citenamefont {De~Lorenzo}(2016)}]{de_lorenzo_optomechanics_2016}%
	\BibitemOpen
	\bibfield  {author} {\bibinfo {author} {\bibfnamefont {L.~A.}\ \bibnamefont
			{De~Lorenzo}},\ }\emph {\bibinfo {title} {Optomechanics with {Superfluid}
			{Helium}-4}},\ \href {https://doi.org/10.7907/Z9RJ4GD7} {Ph.D. thesis},\
	\bibinfo  {school} {California Institute of Technology} (\bibinfo {year}
	{2016})\BibitemShut {NoStop}%
	\bibitem [{\citenamefont {Koolstra}(2019)}]{koolstra_trapping_nodate}%
	\BibitemOpen
	\bibfield  {author} {\bibinfo {author} {\bibfnamefont {G.}~\bibnamefont
			{Koolstra}},\ }\emph {\bibinfo {title} {Trapping a {Single} {Electron} on
			{Superfluid} {Helium} {Using} a {Superconducting} {Resonator}}},\ \href
	{https://doi.org/10.6082/uchicago.2070} {Ph.D. thesis},\ \bibinfo  {school}
	{The University of Chicago} (\bibinfo {year} {2019})\BibitemShut {NoStop}%
	\bibitem [{\citenamefont {Wasserman}\ \emph {et~al.}(2022)\citenamefont
		{Wasserman}, \citenamefont {Harrison}, \citenamefont {Harris}, \citenamefont
		{Sawadsky}, \citenamefont {Sfendla}, \citenamefont {Bowen},\ and\
		\citenamefont {Baker}}]{wasserman_cryogenic_2022}%
	\BibitemOpen
	\bibfield  {author} {\bibinfo {author} {\bibfnamefont {W.~W.}\ \bibnamefont
			{Wasserman}}, \bibinfo {author} {\bibfnamefont {R.~A.}\ \bibnamefont
			{Harrison}}, \bibinfo {author} {\bibfnamefont {G.~I.}\ \bibnamefont
			{Harris}}, \bibinfo {author} {\bibfnamefont {A.}~\bibnamefont {Sawadsky}},
		\bibinfo {author} {\bibfnamefont {Y.~L.}\ \bibnamefont {Sfendla}}, \bibinfo
		{author} {\bibfnamefont {W.~P.}\ \bibnamefont {Bowen}},\ and\ \bibinfo
		{author} {\bibfnamefont {C.~G.}\ \bibnamefont {Baker}},\ }\bibfield  {title}
	{\enquote {\bibinfo {title} {Cryogenic and hermetically sealed packaging of
				photonic chips for optomechanics},}\ }\href
	{https://doi.org/10.1364/OE.463752} {\bibfield  {journal} {\bibinfo
			{journal} {Optics Express}\ }\textbf {\bibinfo {volume} {30}},\ \bibinfo
		{pages} {30822--30831} (\bibinfo {year} {2022})}\BibitemShut {NoStop}%
	\bibitem [{\citenamefont {Powell}, \citenamefont {Roder},\ and\ \citenamefont
		{Hall}(1959)}]{powell_lowtemperature_1959}%
	\BibitemOpen
	\bibfield  {author} {\bibinfo {author} {\bibfnamefont {R.~L.}\ \bibnamefont
			{Powell}}, \bibinfo {author} {\bibfnamefont {H.~M.}\ \bibnamefont {Roder}},\
		and\ \bibinfo {author} {\bibfnamefont {W.~J.}\ \bibnamefont {Hall}},\
	}\bibfield  {title} {\enquote {\bibinfo {title} {Low-temperature transport
				properties of copper and its dilute alloys: Pure copper, annealed and
				cold-drawn},}\ }\href {https://doi.org/10.1103/PhysRev.115.314} {\bibfield
		{journal} {\bibinfo  {journal} {Phys. Rev.}\ }\textbf {\bibinfo {volume}
			{115}},\ \bibinfo {pages} {314--323} (\bibinfo {year} {1959})}\BibitemShut
	{NoStop}%
	\bibitem [{\citenamefont {Pobell}(2007)}]{pobell_matter_2007}%
	\BibitemOpen
	\bibfield  {author} {\bibinfo {author} {\bibfnamefont {F.}~\bibnamefont
			{Pobell}},\ }\href@noop {} {\emph {\bibinfo {title} {Matter and methods at
				low temperatures}}},\ \bibinfo {edition} {3rd}\ ed.\ (\bibinfo  {publisher}
	{Springer},\ \bibinfo {address} {Berlin},\ \bibinfo {year}
	{2007})\BibitemShut {NoStop}%
	\bibitem [{\citenamefont {He}(2020)}]{he_micro_2020}%
	\BibitemOpen
	\bibfield  {author} {\bibinfo {author} {\bibfnamefont {X.}~\bibnamefont
			{He}},\ }\emph {\bibinfo {title} {Micro-resonator optomechanics with
			superfluid helium}},\ \href {https://doi.org/10.14264/uql.2020.123} {\bibinfo
		{type} {{PhD} {Thesis}}},\ \bibinfo  {school} {The University of Queensland}
	(\bibinfo {year} {2020})\BibitemShut {NoStop}%
	\bibitem [{\citenamefont {Abraham}\ and\ \citenamefont
		{Cornell}(1998)}]{abraham_teflon_1998}%
	\BibitemOpen
	\bibfield  {author} {\bibinfo {author} {\bibfnamefont {E.~R.}\ \bibnamefont
			{Abraham}}\ and\ \bibinfo {author} {\bibfnamefont {E.~A.}\ \bibnamefont
			{Cornell}},\ }\bibfield  {title} {\enquote {\bibinfo {title} {Teflon
				feedthrough for coupling optical fibers into ultrahigh vacuum systems},}\
	}\href {https://doi.org/10.1364/AO.37.001762} {\bibfield  {journal} {\bibinfo
			{journal} {Appl. Opt.}\ }\textbf {\bibinfo {volume} {37}},\ \bibinfo {pages}
		{1762--1763} (\bibinfo {year} {1998})}\BibitemShut {NoStop}%
	\bibitem [{\citenamefont {Enss}\ and\ \citenamefont
		{Hunklinger}(2005)}]{enss_low-temperature_2005}%
	\BibitemOpen
	\bibfield  {author} {\bibinfo {author} {\bibfnamefont {C.}~\bibnamefont
			{Enss}}\ and\ \bibinfo {author} {\bibfnamefont {S.}~\bibnamefont
			{Hunklinger}},\ }\href@noop {} {\emph {\bibinfo {title} {Low-temperature
				physics}}}\ (\bibinfo  {publisher} {Springer Science \& Business Media},\
	\bibinfo {year} {2005})\BibitemShut {NoStop}%
	\bibitem [{\citenamefont {Sabisky}\ and\ \citenamefont
		{Anderson}(1973)}]{sabisky_verification_1973}%
	\BibitemOpen
	\bibfield  {author} {\bibinfo {author} {\bibfnamefont {E.~S.}\ \bibnamefont
			{Sabisky}}\ and\ \bibinfo {author} {\bibfnamefont {C.~H.}\ \bibnamefont
			{Anderson}},\ }\bibfield  {title} {\enquote {\bibinfo {title} {Verification
				of the lifshitz theory of the van der waals potential using liquid-helium
				films},}\ }\href {https://doi.org/10.1103/PhysRevA.7.790} {\bibfield
		{journal} {\bibinfo  {journal} {Phys. Rev. A}\ }\textbf {\bibinfo {volume}
			{7}},\ \bibinfo {pages} {790--806} (\bibinfo {year} {1973})}\BibitemShut
	{NoStop}%
	\bibitem [{\citenamefont {Clements}\ \emph {et~al.}(1993)\citenamefont
		{Clements}, \citenamefont {Epstein}, \citenamefont {Krotscheck},\ and\
		\citenamefont {Saarela}}]{clements_structure_1993}%
	\BibitemOpen
	\bibfield  {author} {\bibinfo {author} {\bibfnamefont {B.~E.}\ \bibnamefont
			{Clements}}, \bibinfo {author} {\bibfnamefont {J.~L.}\ \bibnamefont
			{Epstein}}, \bibinfo {author} {\bibfnamefont {E.}~\bibnamefont
			{Krotscheck}},\ and\ \bibinfo {author} {\bibfnamefont {M.}~\bibnamefont
			{Saarela}},\ }\bibfield  {title} {\enquote {\bibinfo {title} {Structure of
				boson quantum films},}\ }\href {https://doi.org/10.1103/PhysRevB.48.7450}
	{\bibfield  {journal} {\bibinfo  {journal} {Phys. Rev. B}\ }\textbf {\bibinfo
			{volume} {48}},\ \bibinfo {pages} {7450--7470} (\bibinfo {year}
		{1993})}\BibitemShut {NoStop}%
	\bibitem [{\citenamefont {Clements}, \citenamefont {Krotscheck},\ and\
		\citenamefont {Lauter}(1993)}]{clements_growth_1993}%
	\BibitemOpen
	\bibfield  {author} {\bibinfo {author} {\bibfnamefont {B.~E.}\ \bibnamefont
			{Clements}}, \bibinfo {author} {\bibfnamefont {E.}~\bibnamefont
			{Krotscheck}},\ and\ \bibinfo {author} {\bibfnamefont {H.~J.}\ \bibnamefont
			{Lauter}},\ }\bibfield  {title} {\enquote {\bibinfo {title} {Growth
				instability in helium films},}\ }\href
	{https://doi.org/10.1103/PhysRevLett.70.1287} {\bibfield  {journal} {\bibinfo
			{journal} {Phys. Rev. Lett.}\ }\textbf {\bibinfo {volume} {70}},\ \bibinfo
		{pages} {1287--1290} (\bibinfo {year} {1993})}\BibitemShut {NoStop}%
	\bibitem [{\citenamefont {Kashkanova}\ \emph
		{et~al.}(2017{\natexlab{b}})\citenamefont {Kashkanova}, \citenamefont
		{Shkarin}, \citenamefont {Brown}, \citenamefont {Flowers-Jacobs},
		\citenamefont {Childress}, \citenamefont {Hoch}, \citenamefont {Hohmann},
		\citenamefont {Ott}, \citenamefont {Reichel},\ and\ \citenamefont
		{Harris}}]{kashkanova_optomechanics_2017}%
	\BibitemOpen
	\bibfield  {author} {\bibinfo {author} {\bibfnamefont {A.~D.}\ \bibnamefont
			{Kashkanova}}, \bibinfo {author} {\bibfnamefont {A.~B.}\ \bibnamefont
			{Shkarin}}, \bibinfo {author} {\bibfnamefont {C.~D.}\ \bibnamefont {Brown}},
		\bibinfo {author} {\bibfnamefont {N.~E.}\ \bibnamefont {Flowers-Jacobs}},
		\bibinfo {author} {\bibfnamefont {L.}~\bibnamefont {Childress}}, \bibinfo
		{author} {\bibfnamefont {S.~W.}\ \bibnamefont {Hoch}}, \bibinfo {author}
		{\bibfnamefont {L.}~\bibnamefont {Hohmann}}, \bibinfo {author} {\bibfnamefont
			{K.}~\bibnamefont {Ott}}, \bibinfo {author} {\bibfnamefont {J.}~\bibnamefont
			{Reichel}},\ and\ \bibinfo {author} {\bibfnamefont {J.~G.~E.}\ \bibnamefont
			{Harris}},\ }\bibfield  {title} {\enquote {\bibinfo {title} {Optomechanics in
				superfluid helium coupled to a fiber-based cavity},}\ }\href
	{https://doi.org/10.1088/2040-8986/aa551e} {\bibfield  {journal} {\bibinfo
			{journal} {J. Opt.}\ }\textbf {\bibinfo {volume} {19}},\ \bibinfo {pages}
		{034001} (\bibinfo {year} {2017}{\natexlab{b}})}\BibitemShut {NoStop}%
	\bibitem [{\citenamefont {Kurihara}(1981)}]{kurihara_large-amplitude_1981}%
	\BibitemOpen
	\bibfield  {author} {\bibinfo {author} {\bibfnamefont {S.}~\bibnamefont
			{Kurihara}},\ }\bibfield  {title} {\enquote {\bibinfo {title}
			{Large-{Amplitude} {Quasi}-{Solitons} in {Superfluid} {Films}},}\ }\href
	{https://doi.org/10.1143/JPSJ.50.3262} {\bibfield  {journal} {\bibinfo
			{journal} {J. Physical Soc. Japan}\ }\textbf {\bibinfo {volume} {50}},\
		\bibinfo {pages} {3262--3267} (\bibinfo {year} {1981})}\BibitemShut {NoStop}%
\end{thebibliography}
\end{document}